\documentclass[preprint,12pt]{elsarticle}

\usepackage{amssymb}
\usepackage{rotating}

\journal{Parallel Computing}

\begin{document}

\begin{frontmatter}



\title{Coalesced communication: a design pattern for complex parallel scientific software}


\author[complex,ucl]{Hywel B. Carver}
\author[ucl]{Derek Groen}
\author[ucl]{James Hetherington}
\author[ucl]{Rupert W. Nash} 
\author[ucl,complex]{Miguel O. Bernabeu}
\author[ucl]{Peter V. Coveney}

\address[complex]{CoMPLEX, University College London, London, United Kingdom}
\address[ucl]{Centre for Computational Science, University College London, London, United Kingdom\fnref{emailref}}

\fntext[emailref]{E-mail: hywel.carver.09@ucl.ac.uk (Hywel B. Carver), p.v.coveney@ucl.ac.uk (Peter V. Coveney)}

\begin{abstract}
We present a new design pattern for high-performance parallel scientific
software, named coalesced communication. This pattern allows for a structured
way to improve the communication performance through coalescence of multiple
communication needs using two communication management components. We apply the
design pattern to several simulations of a lattice-Boltzmann blood flow solver
with streaming visualisation which engenders a reduction in the communication
overhead of approximately 40\%. 
\end{abstract}


\end{frontmatter}


\section{Introduction}
 
High-performance parallel scientific software often consists of
complex, multi-functional, multi-physics software components, run on
infrastructures which are increasingly large and frequently hybrid in nature
(e.g., featuring many-core architectures or distributed systems).
Orchestrating the work of these components requires advanced software
engineering and design approaches to manage the attendant complexity. The
result is that the structure of high-performance computing codes is moving
towards the use of higher-level design abstractions.
One way to capture these design abstractions is through the definition of {\em
design patterns}. Design patterns are commonly applied in software
engineering~\cite{Erich:1995}. They are formal definitions which describe a
specific solution to a design problem, and can be found in a range of
scientific and engineering disciplines. With high performance computing (HPC) codes
growing in complexity, existing design patterns are more commonly applied in HPC
and numerous new design patterns have emerged~\cite{Ortega-Arjona:2010,Mattson:2004}.

Here we present a new design pattern: coalesced communication. In this pattern,
each component registers the communication tasks it will require during the
different stages, or {\em steps}, of execution with a central registry. We
refer to each component which wishes to register communication requests as a
{\em Client}. This registry analyses the required communications and combines
requests from each Client at appropriate steps of the execution. This allows
work of one Client (such as a scientific kernel) to overlap with the
communication of another Client (such as streaming visualisation or error
correction), and results in a single synchronization point between processes
during each step.

Several groups have experimented with the coalescence of communication,
although none of these have developed this into a generalised design
pattern.  {\em Bae et al.}~\cite{Bae96} benchmark the coalescence of communication as
a factor influencing code complexity and efficiency within two algorithms. {\em Bell
et al.}~\cite{Bell:2003} investigate the performance benefit of overlapping
communication with communication, which is an alternative method to reduce the
number of synchronisation points. {\em Chavarria et al.}~\cite{Chavarria05} implement
a form of coalescence in a High-Performance Fortran compiler for situations
where one code location has multiple communication events, and find a reduction
of up to 55\% in communication volume. {\em Chen et al.}~\cite{Chen05} find similar
performance improvement when applying coalescing in programs written in Unified
Parallel C, and {\em Koop et al.}~\cite{Koop07} report significant improvements
in throughput when using low-level coalescence for sending small MPI messages.

\section{Coalesced communication}

The coalesced communication pattern is applicable to any parallel software
which carries out multiple tasks, and therefore has a range of communication
needs. These communication needs may, for example, include exchanges required for
one or more scientific kernels, visualisation, steering, dynamic domain
decomposition, coupling with one or more external programs, introspection or
error recovery. Of course, each of these Clients could do its own
communication internally, but this can be highly inefficient from a performance
perspective due to the large number of synchronisation points with other
processes. The coalesced communication pattern allows us to improve the
communication performance by reducing the number of synchronisation points in
an organised way. 

\begin{figure}[!t]
\centering
\includegraphics[width=4.5in]{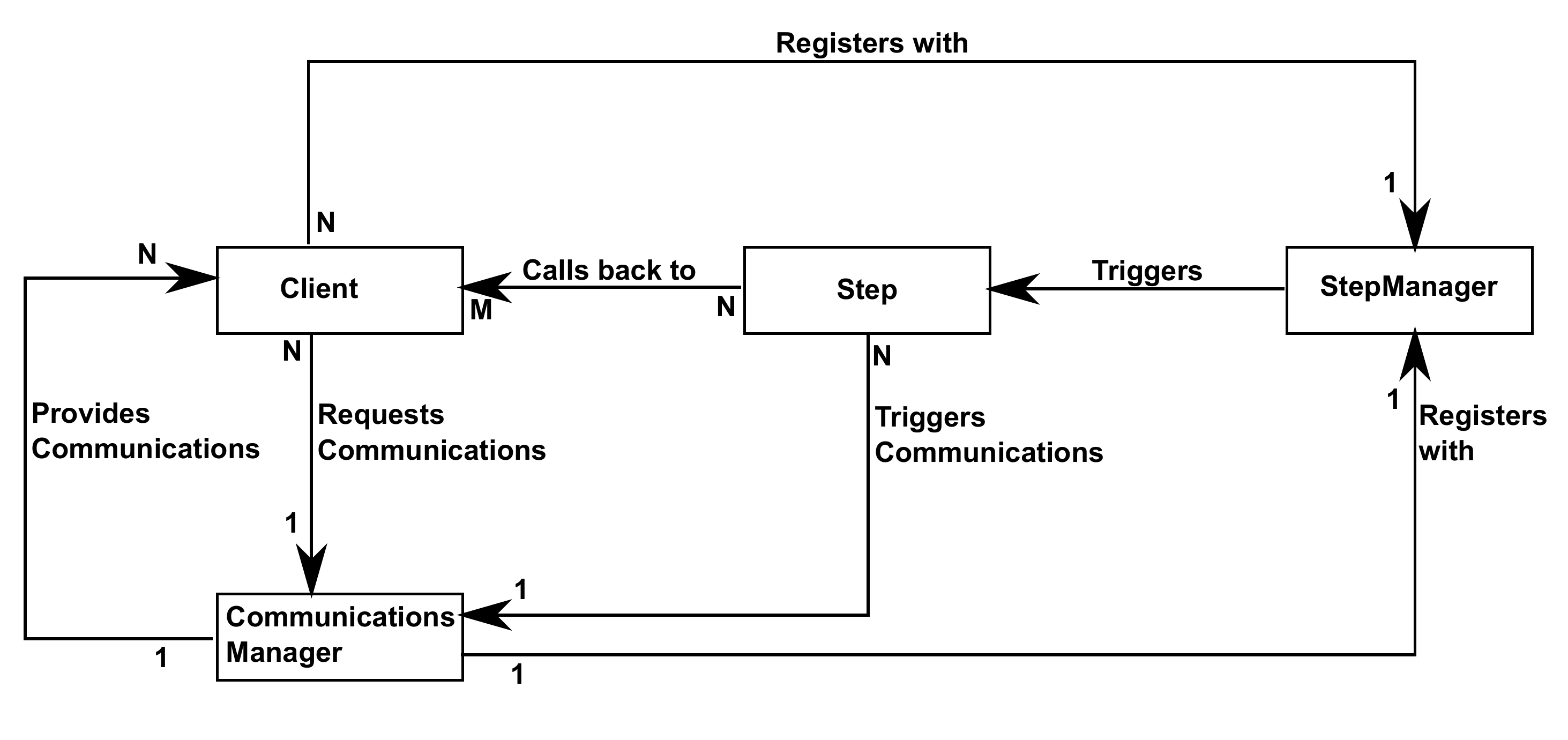}
\caption{Entity relationship diagram of the coalesced communication design pattern.}
\label{Fig:erd}
\end{figure}

Within the coalesced communication pattern, each Client registers with an
administrative object called the {\em StepManager}, and all communication is
indirected through a central store of communication requirements called the
{\em CommunicationsManager} object. The relations of these objects are shown in
Figure~\ref{Fig:erd}.  In each of several {\em Steps}, a call back is made to
each Client to carry out those computations that are safe to perform during
that step, while the CommunicationsManager object makes the appropriate MPI
calls to initiate non-blocking message passing for each requested piece of
communications. In this way, the communications of all Clients can be
overlapped with their calculation, potentially providing substantial
performance gains. In addition, the bundling of all the non-blocking
communications reduces the number of synchonisation points here to one. 

\begin{figure}[!t]
\centering
\includegraphics[width=5.2in]{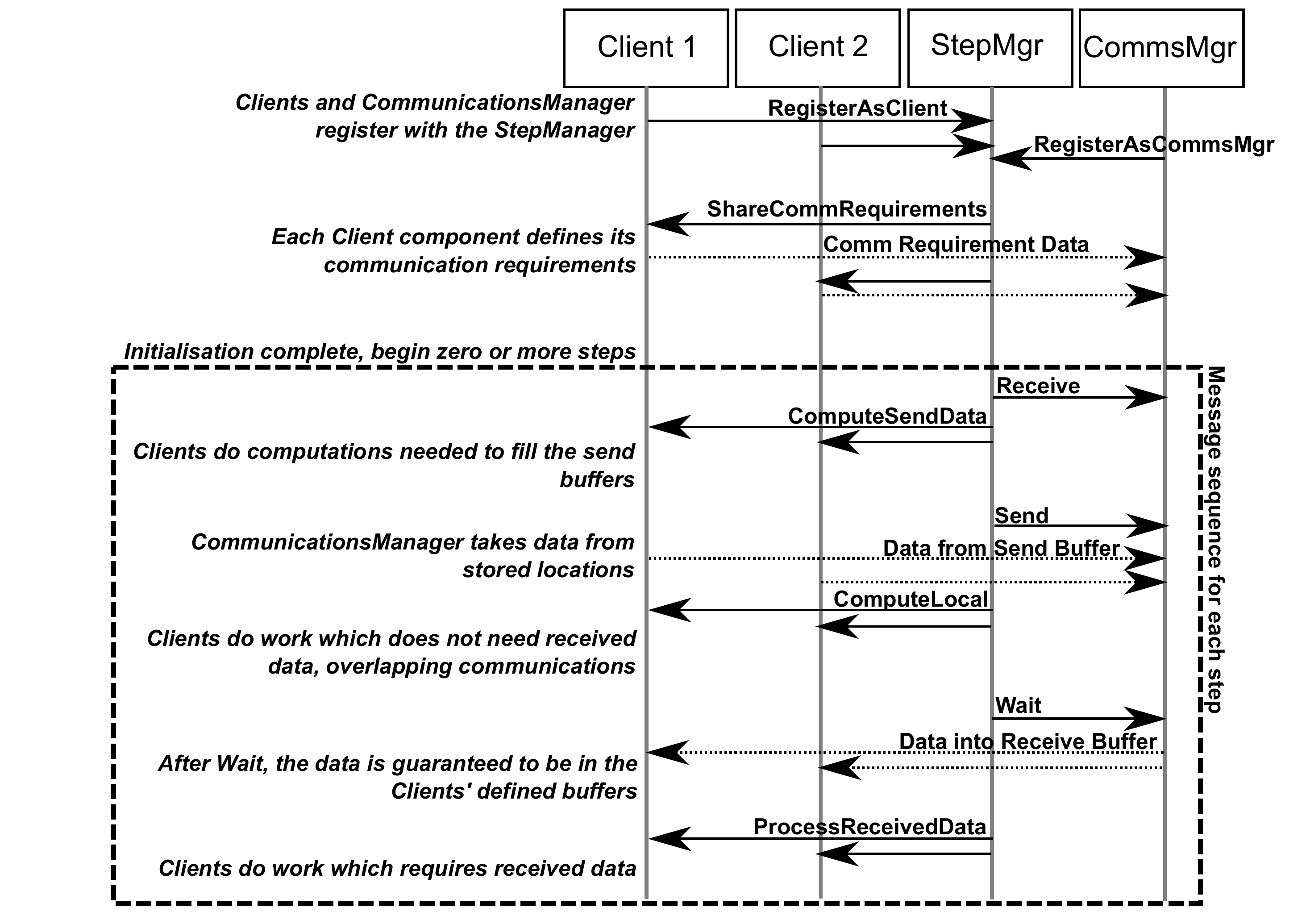}
\caption{Message sequence chart of the coalesced communication pattern,
generalized for an application with two Client components which require 
communications. Function calls and data movements are indicated respectively 
with solid and dashed arrows. The StepManager and CommunicationsManager 
objects are abbreviated respectively as StepMgr and CommsMgr. Time 
proceeds vertically downwards.}
\label{Fig:msc}
\end{figure}

We present the sequence of events for an application with two Clients in
Figure~\ref{Fig:msc}. Here we see computation callbacks preceding and following
each of the MPI send, receive, and wait calls. For example, computation
callbacks are made to each Client after the CommunicationsManager makes the MPI
send calls, while it waits to receive the incoming data. The incoming data are
placed into buffers registered with the CommunicationsManager at the beginning
of each step, but the data is only safe to use following completion of the Wait
call made by CommunicationsManager.

\section{Implementation}

We have implemented the coalesced communication design pattern within the
HemeLB lattice-Boltzmann simulation environment, which is intended to
accurately model cerebrovascular blood flow. HemeLB is written in C++ and aims
to provide timely and clinically relevant assistance to
neurosurgeons~\cite{Mazzeo:2008}. HemeLB contains a range of functionalities,
including the core lattice-Boltzmann kernel, visualisation modules and a
steering component which allows for interactive use of the application. HemeLB
has been shown to efficiently model sparse geometries using up to at least
32,768 compute cores~\cite{Groen:2012}; {\em inter alia}, has been used for a
variety of scenarios~\cite{Carver:2012,Mazzeo:2008}.

The primary Clients registered with the StepManager within HemeLB are those
raised by the core lattice-Boltzmann kernel, an {\em in situ} visualisation
module and an module for introspective monitoring. However, HemeLB will frequently
run with additional Clients as there are a number of optional modules, such as the
computational steering server. Within this article we focus on only the core 
lattice-Boltzmann communications and the visualisation communications.

%

\section{Performance Tests}

We have run HemeLB on 1024 cores on the HECToR Cray XE6 machine in Edinburgh,
United Kingdom, using a sparse cerebrovascular bifurcation simulation domain
which contains 19,808,107 fluid sites. Our simulations run for 2000 steps with
three different settings, rendering respectively 10, 100 and 200
images using the visualisation module. We repeated each run both with and 
without coalesced communication enabled, using a
compile-time parameter to toggle this functionality. We measured the total
time spent on the simulation, on all communications, and on local operations 
required for constructing the images. 

We present the results of our performance tests in Table~\ref{Tab:CCperf}.
Based on our measurements we find that the communication overhead in our
coalesced runs amounts to between 57 and 63\% of the overhead in the
non-coalesced runs. When we render more images per timestep, the absolute
performance benefit increases while relative performance benefit slightly
decreases. However, the frame rate we obtain for the runs with 200 images
generated is already sufficient for real-time visual inspection of the data.
The time spent on visualisation is 0.0034 second per image, and scales 
linearly with the number of images rendered.

\begin{table}[!t] 

\caption{Performance results of our HemeLB simulations, run with and without
the coalesced communication strategy. Each simulation ran for 2000 time steps,
using 1024 cores and modelling blood flow in a bifurcation simulation domain. 
We ran our simulations rendering respectively 10 images (first two rows), 100 
images (middle two rows), and 200 images (last two rows) at evenly spaced time 
intervals during execution.}
\label{Tab:CCperf}

\centering 
\begin{tabular}{lllll} 
\hline 
\# of images & Coalesced Comm. & Total time & Comm. time & Vis. time\\
&  & [s] & [s] & [s]\\
\hline
10  & enabled  & 27.6 & 2.36 & 0.03\\
10  & disabled & 29.3 & 4.07 & 0.03\\
\hline
100 & enabled  & 30.0 & 3.13 & 0.34\\
100 & disabled & 31.9 & 5.15 & 0.33\\
\hline
200 & enabled  & 32.7 & 3.82 & 0.68\\
200 & disabled & 34.8 & 6.07 & 0.66\\
\hline 
\end{tabular} 
\end{table}

\section{Discussion and conclusions}

We have presented the coalesced communication design pattern, which allows the
coalescence of the interprocess communications of multiple Client components
within complex parallel scientific software. We have demonstrated the benefit
of adopting the design pattern based on an implementation in a blood flow
application. Here the use of coalesced communication reduces the total
communication overhead of the simulations, which have two primary Clients, by
approximately 40\%. This improvement results in the application taking about
7\% less time overall, making it more responsive when applied for clinical or
scientific purposes. The design pattern can be directly applied in other
parallel scientific software projects, allowing for a structured way to improve
the communication performance through coalescence. 

\section{Acknowledgements}

This work has received funding from the CRESTA and MAPPER projects within the
EC-FP7 (ICT-2011.9.13) under Grant Agreements nos. 287703 and 261507, the
British Heart Foundation, and from EPSRC Grants EP/I017909/1
(www.2020science.net) and EP/I034602/1. This work made use of the HECToR
supercomputer at EPCC in Edinburgh, funded by the Office of Science and
Technology through EPSRC's High End Computing Programme.


\bibliographystyle{elsarticle-num}
\bibliography{coalesced.bib}







\end{document}